\documentclass[aps,twocolumn,amsmath,amssymb,floatfix,superscriptaddress,prb]{revtex4-2}

\usepackage{dcolumn}
\usepackage{bm,graphicx,url,epsf,color}
\usepackage{amssymb}
\usepackage{amsthm}
\usepackage{comment}
\usepackage{physics}
\usepackage[colorlinks=true,linkcolor=blue,citecolor=blue,linktocpage=true,breaklinks=true]{hyperref}
\usepackage{cleveref}
\usepackage{soul}

\newcommand{\bk}{\bm{k}}
\newcommand{\br}{\bm{r}}
\newcommand{\bq}{\bm{q}}

\newcommand{\bR}{\bm{R}}

\newcommand{\cinam}{CNRS/Aix-Marseille Université, Centre Interdisciplinaire de Nanoscience de Marseille UMR 7325 Campus de Luminy, 13288 Marseille cedex 9, France}

\begin{document}

\title{Moir\'e amplification of highly tunable shift current response in twisted trilayer graphene}
\author{Yuncheng Mao}
\author{Claudio Attaccalite}
\author{Diego Garc\'ia Ovalle}
\affiliation{\cinam}
\date{\today}

\begin{abstract}
{
In this work we analyze the shift current conductivity in helical twisted trilayer graphene. Without loss of generality, we show that the density of states and the twist angle set an upper bound for this response, which is inversely proportional to the square of the twist angle. For the case of ABA stacking and at the magic angle, the shift photoconductivity can reach values of order $10^4~\mathrm{\mu A \cdot nm \cdot V}^{-2}$ for frequencies below 50 meV, which can be attributed to the interband transitions between the two flattened middle bands close to the Fermi level. By tuning the twist angle, we demonstrate that the photoconductivity is shifted in the frequency range and it is further influenced by two additional factors: The magnitude of the quantum geometric measure and the energy separation between the bands. Furthermore, we propose a scenario in the AAA stacked configuration, where the photoconductivity can be of order $10^5~\mathrm{\mu A \cdot nm \cdot V}^{-2}$ in the THz regime, revealing a potential influence of the stacking in the optimization of the shift current conductivity. Therefore, a large density of states, a small twist angle and the layer stacking are ingredients that hold promising functionality for photovoltaic applications in moir\'e heterostructures.}

\end{abstract}

\maketitle

\section{Introduction}

Nowadays, solar-cell technology is a suitable option for high efficient and eco-friendly energy applications. In this context, an alternative mechanism to the standard PN photovoltaics for the  photocurrent generation is the bulk photovoltaic effect (BPVE). In BPVE the photocurrent generation is driven by the shift current (SC), where light illumination induces a photoconductivity in materials lacking inversion symmetry \cite{Kraut1979,belinicher1980photogalvanic,Baltz1981}. 
Its origin has been attributed to the shift of the center of charges in the presence of light. In contrast to the conventional PN photovoltaic effect, the BPVE can generate a photovoltage above the band gap, allows for overcoming the Shockley-Queisser limit and in principle surpassing the conversion efficiency of conventional solar cells. 
However, even with all these good premises, BPVE is still not competitive with standard PN Photovoltaics because finding a system with a large average shift vector upon absorption of a photon is extremely complicated \cite{Pusch2023}, suggesting further exploration on materials with different band structure properties \cite{cook2017design}. 
Concerning nonmagnetic materials, natural prospects for large shift conductivities are ferroelectrics, such as BaTiO$_3$ or PbTiO$_3$ \cite{Young2012}, but its dependence with the direction and strength of the ferroelectric polarization is not fully understood.
On the other hand, since the shift photoconductivity only requires inversion symmetry breaking, there is a growing interest on the study of BPVE in metallic and semi-metallic systems as potential substitutes to photovoltaics in narrow-gap semiconductors, where the SC can be enhanced by real and virtual transitions between the bands, in materials like the topological semimetal TaAs \cite{Zhang2018,Osterhoudt2019}. 
In fact, when the frequency of the incoming light lies within the mid-infrared region of the spectrum, it can exceed by orders of magnitude the photoconductivity of conventional ferroelectrics \cite{Young2012,Tiwari2022} and chiral materials \cite{Zhang2019}. Nevertheless, it is still smaller than the gigantic SC response computed in BiTeI reaching about 750~$\mathrm{\mu A \cdot V^{-2}}$~\cite{Xu2020}, which can further be tuned through a topological transition in the presence of pressure~\cite{Tan2016}.

Recently, a wide variety of investigations have revealed that twisted multilayer systems made of transition metal dichalcogenides (TMD) or graphene monolayers, just to name a few, are able to display colossal nonlinear d.c. responses \cite{He2021b,Pantaleon2021,Duan2022,Huang2022,Sinha2022,Zhang2022,Huang2023} and second order photocurrents \cite{Liu2020,Xu2021,chaudhary2022shift,Kaplan2022,Zheng2023,Silvanes2023,Hu2023,Postlewaite2024,Shayeganfar2024}. 
Regarding the nonlinear Hall effect, the magnitude of the conductivity in twisted bilayer WTe$_2$ \cite{He2021b} relies on the numerous band crossings and band inversions around the Fermi level, leading to an enhancement of the Berry curvature. Correlation effects have also been proposed as an explanation to the large nonlinear Hall effect in WSe$_2$ \cite{Huang2022}. On the other hand, graphene-type materials usually exhibit rotational symmetries, typically $\textrm{C}_3$ or $\textrm{C}_6$, forbidding the second order Hall effect driven by the Berry curvature dipole \cite{Du2021,Ovalle2024} and the intrinsic nonlinear Hall effect \cite{Gao2023}.
In this situation, it is more likely that the d.c. current could be associated with more complex extrinsic contributions such as side jump or skew scattering \cite{He2021}. Although the rotational symmetry can be broken in several ways, e.g., by applying strain \cite{Son2019,Chakraborty2022} or modifying the structural phase of the material \cite{Lesne2023}, the nonlinear photocurrents can be considered as a signal that is more flexible to the crystal symmetry constraints, and a versatile route to take advantage of the electronic properties of moir\'e patterns. Several features of twisted materials can strongly impact the overall magnitude of the shift conductivity: For twisted bilayer graphene (TBG) in the THz regime, it can reach $10^3~\mathrm{\mu A \cdot nm \cdot V^{-2}}$ depending on the value of the applied strain \cite{Kaplan2022}. This value is sensitive to the presence of particle hole symmetry, encapsulation with hBN and perturbative effects from a substrate \cite{Penaranda2024}. When the number of layers is more than $5$, the shift conductivity can increase up to $10^4~\mathrm{\mu A \cdot nm \cdot V^{-2}}$, whose origin relies on the virtual transitions between the multiple flat bands involved \cite{Chen2024}. Motivated by the facts mentioned above, and the rebirth of the interest in semi-direct materials to observe maximal SC, we study the origin, intensity and the potential tunability of this signal in helical twisted trilayer graphene (hTTG).

Our paper is organized as follows: In Section~\ref{theory} we show analytically that an upper bound of the SC response is given by the product of the density of states involving two bands, whose distance corresponds to the energy of the incident photon. This upper bound is a complement to a previous upper limit reported for the shift conductivity in extended systems \cite{Tan2019}. 
In this section we also present the model for hTTG, which we later use in Section~\ref{sec results} to provide the numerical results and a discussion about the outcomes of the work. Finally, in Section~\ref{conclusion} we state the main conclusions of the article.

\section{Theory and model}\label{theory}

In this section, we first scrutinize how the upper bound of SC response is set, and then explicitly shown amplification factor proportional to $1/\theta^2$, which is specific to such twisted moir\'e materials.
We will take into account these observations in the case of hTTG, which is defined in the next section. This is followed by our numerical calculations of the SC conductivity and a proper discussion about our results.
We remark that the shift photoconductivities in the THz regime estimated by the single-particle description is not expected to be modified substantially in the presence of electron interactions considered up to Hartree-Fock level~\cite{chaudhary2022shift}. But for higher frequency light, the results from Ref~~\cite{chaudhary2022shift} imply that interaction (up to Hartree-Fock level) may additionally amplify the SC response, which is not investigated in the single-particle picture. 

\subsection{Upper bound of the shift current conductivity}
First, we show that the shift current response is bounded by the density of states. The second harmonic generation current reads $\textrm{J}^{a}=\sigma^{a;bc}(0, \omega, -\omega)\mathcal{E}^b(\omega)\mathcal{E}^c(-\omega)$, where $\mathcal{E}(\omega)$ is an AC electric field. Upon linearly polarized light and when time reversal symmetry is preserved, the SC conductivity for a periodic Hamiltonian that satisfies the Schrodinger equation $H(\bk)\ket{u_m} = E_{n\bk} \ket{u_m}$ is given by \cite{Sipe2000}:
\begin{equation}\label{shift current}
\begin{split}
\sigma^{a;bc}(0, \omega, -\omega) = - \frac{i \pi e^3}{\hbar^2} \sum_{n,m} \int \frac{\dd^2\bk}{(2\pi)^2}~ \delta(\omega-\omega_{mn}) \times \\ 
f_{nm} (r^b_{mn} r^c_{nm;a} + r^c_{mn}r^b_{nm;a}),
\end{split}
\end{equation}
where the $\bk$-dependence of $f_{nm}$, $r^b_{mn}$, $r^c_{nm;a}$ and $\omega_{mn}$ is implicit. The subscripts $n$ and $m$ go through the band indices and $a, b, c$ $\in \{x, y\}$ denote the directions of the output current and the applied electric field. Besides, $f_{nm} = f(E_{n\bk}) - f(E_{m\bk})$, being $f(\mathcal{E})$ the Fermi-Dirac distribution, $\mu$ the chemical potential, $k_B$ is the Boltzmann constant and $T$ is the temperature of the system. Throughout this paper, we will assume that $\mu = 0$ and $T = 0$, unless stated otherwise explicitly. 
In addition, $\omega_{mn} = (E_{m\bk} - E_{n\bk})/\hbar$ and 
$r^a_{nm} = i \mel{u_n}{\partial_{k_a}}{ u_m}$ for $n \neq m$ is the interband Berry connection, while $r^b_{nm;a}$ is the generalized derivative of $r^b_{nm}$ with respect to $k_a$ given by:
\begin{equation}\label{eq: gdev r}
    r^b_{nm;a} = \left[\partial_{k_a} -i (\mathcal{A}^a_{nn} -\mathcal{A}^a_{mm})\right] r^b_{nm},
\end{equation}
where $\mathcal{A}^a_{nn}$ is the component of the \emph{intraband} Berry connection  $\bm{A}_{nn}$ along the direction $a$ and $\bm{A}_{nn} = i \mel{u_{n\bk}}{\bm \nabla_{\bk}}{u_{n\bk}}$.
Although Eq.~\eqref{shift current} is gauge invariant, the numerical calculation of $\sigma^{a;bc}$ is greatly complicated by explicit gauge choice at every $\bk$ point for the estimation of derivatives with respect to $\bk$. Then, it is preferable to employ the gauge-free expressions for the interband connections
\begin{equation}\label{eq: connection}
    r^a_{nm} = \frac{-i \mel{u_{n\bk}}{\partial_{k_a}H}{u_{m\bk}}}{E_{n\bk}-E_{m\bk}} = \frac{v^a_{nm}}{i \omega_{nm}}, ~ n \neq m,
\end{equation}
 Using $v^a_{nm} = \mel{u_{n\bk}}{\partial_{k_a}H}{u_{m\bk}}/\hbar = v_F \mel{u_{n\bk}}{\sigma_a}{u_{m\bk}}$, $r^b_{nm;a}$ can be conveniently computed with the gauge-free expression\cite{cook2017design}
\begin{equation}\label{eq: gderiv connection}
\begin{split}
    r^{b}_{nm;a} = \frac{i}{\omega_{nm}}\sum_{l\neq n,m}\left( \frac{v^b_{nl} v^a_{lm}}{\omega_{lm}}-\frac{v^a_{nl}v^b_{lm}}{\omega_{nl}} \right) \\
    + \frac{i}{\omega^2_{nm}}\left( v^a_{nm} \Delta^b_{nm} + v^b_{nm} \Delta^a_{nm} \right), ~n \neq m .
\end{split}
\end{equation}
Notice that each of the $r^a_{nm}$ and $r^b_{nm;a}$ alone is not gauge invariant. They acquire a phase factor $e^{-i(\phi_n(\bk) - \phi_m(\bk))}$ upon a gauge change $\ket{u_n} \rightarrow \ket{u_{n\bk}}e^{i\phi_n(\bk)}$. However, the product $r^b_{mn} r^c_{nm;a}$ cancels out the phase brought by gauge change, and is thus gauge-invariant. In the following, $\sigma^{a;bc}(\omega)$ is used to represent $\sigma^{a;bc}(0, \omega, -\omega)$ without ambiguity to simplify the notation. Let us denote $X^{a;bc}_{nm} (\bk) = f_{nm} (r^b_{mn} r^c_{nm;a} + r^c_{mn}r^b_{nm;a})$. By applying the Cauchy-Schwartz inequality, we have
\begin{equation}
\begin{split}    
    & \left| \int\frac{\dd\bk}{(2\pi)^2} X^{a;bc}_{nm}(\bk) \delta(\omega - \omega_{mn}) \right |^2 \leqslant \\
    & \left|\int\frac{\dd\bk}{(2\pi)^2} X^{a;bc}_{nm}(\bk) \right|^2 \cdot \left|\int\frac{\dd\bk}{(2\pi)^2} \delta(\omega - \omega_{mn})\right|^2 .
\end{split}
\end{equation}
\noindent 
The second integral on the right-hand side can be regarded as the joint density of states (JDOS) at zero temperature, assuming that the band $n$ is filled, while band $m$ is empty. This defines the upper limit of the SC conductivity. Although JDOS is often discussed in the literature studying SC response~\cite{yang2018divergent}, its physical picture is not straightforwardly understandable. In fact, it can be further shown that the upper bound of SC conductivity can be given by the product of the density of states at two energy levels separated by the photon energy. Using H\"older's inequality, one obtains

\begin{equation}\label{sc upper lim}
\begin{split}
    &|\sigma^{a;bc}(\omega)| \leqslant \iint \dd \omega_1\dd \omega_2 \sum_{nm} \int\frac{\dd^2 \bk}{(2\pi)^2} \times \\
    & \left|\frac{\pi e^3}{\hbar^2} X^{a;bc}_{nm}(\bk)\right| \delta(\omega-\omega_1+\omega_2) \times \rho(\omega_1) \rho(\omega_2) .
\end{split}
\end{equation}

The physical interpretation of Eq.~\eqref{sc upper lim} is therefore clear. $\omega_1$ and $\omega_2$ designate an energy level of the conduction bands and another energy level in the valence bands, respectively. For a photon of energy $\omega = \omega_1 - \omega_2$, the maximum SC response is bounded by the product of the density of states at both energy levels at $\omega_1$ and $\omega_2$. Eq.~\eqref{sc upper lim} strongly implies that the high density of states brought by the flattened bands in moir\'e materials, such as TBG and TTG, can potentially enhance the SC response. 
In Appendix~\ref{sc in graphene} we use a toy model of a gapped graphene single layer to show that the high DOS peaks are indeed the decisive factor in the SC response for these type of materials. 

\subsection{Moir\'e amplification of shift current}\label{sec: sc amplification}
In twisted multilayer graphene systems, the typical length is set by $1/q$, being $q$ the nearest distance between the $\mathbf{K}$ points of adjacent graphene layers in reciprocal space. In twisted bilayer graphene with a twist angle $\theta$, we have $q = |\mathbf{K}| \theta$. The typical energy scale is then set by $\hbar v_F q = \theta \times 9.905~\mathrm{eV}$, if the convention $\hbar v_F |\mathbf{K}| = 9.905~\mathrm{eV}$ is applied. Here, $v_F$ is the Fermi velocity of a graphene single layer.
Expressing the SC conductivity in dimensionless units, we deduce that 
\begin{equation}
    \sigma^{a;bc}(\omega) = \frac{\sigma_0}{\theta^2} \Theta^{a;bc}(\omega), 
\end{equation}
where
\begin{equation}\label{typical sc cond}
    \sigma_0 = \frac{\pi e^2}{\hbar^2 v_F |\mathbf{K}|^2} = 4.52~\mathrm{\mu A \cdot nm \cdot V^{-2}} 
\end{equation}
is a constant quantity entirely defined by the parameters of a single layer graphene.
Here we adopt the parameters $\hbar v_F |\mathbf{K}| = 9.905~\mathrm{eV}$, and consider the C-C bond length in graphene to be $1.42$ \AA. $\Theta^{a;bc}(\omega)$ is a dimensionless quantity that does not depend explicitly on the twist angle:
\begin{equation}
\begin{split}
    \Theta^{a;bc}(\omega) =& -i \sum_{n,m} \int \frac{\dd^2(\bk/q^2)}{(2\pi)^2} \delta\left(\frac{\omega - \omega_{mn}}{v_F q}\right) \times \\  
    & q^3 f_{nm} (r^b_{mn} r^c_{nm;a} + r^c_{mn}r^b_{nm;a}) .
\end{split}
\end{equation}
Considering the magic angle of TBG to be $\theta = 1.09^\circ$, we have 
\begin{equation}
    \sigma_0/\theta^2 = 1.25 \times 10^4 ~\mathrm{\mu A \cdot nm \cdot V^{-2}}~,
\end{equation}
which strongly suggests the potential of twisted moir\'e materials in amplifying the shift photoconductivity, with the SC conductivity being proportional to $1/\theta^2$. 
Therefore, the small rotation angle of twisted bilayer or trilayer graphene controls the ``amplification'' factor of the SC response in such moir\'e materials.

Notice that it should \emph{not} be understood that the SC conductivity diverges at $\theta \rightarrow 0$. Although the Bistritzer-MacDonald (BM) continuum model \cite{bistrizer2011moire} of TBG describes the physics at small twist angle below $10^\circ$, it is still under debate what is the lower limit of the twist angle where the BM model remains valid\cite{PhysRevB.107.155403}. As a consequence, TBG and TTG near magic angles are potent candidates for SC generation, where both of the small twist angle and the high density of states brought by flattened bands are in favor of significant SC response. As we will show later, the SC conductivity in TTG is indeed several orders of magnitude greater than that of a gapped single layer graphene.

\subsection{Quantum geometry in shift current response}
It is also clear that the dimensionless quantity $\Theta^{a;bc}(\omega)$ incorporates all the contribution from the quantum geometry and (joint) density of states of the wave functions.
More specifically, the dimensionless quantity 
\begin{equation}\label{eq: dimless measure}
    \tilde{X}^{a;bc}_{nm}(\bk) = q^3 (r^b_{mn} r^c_{nm;a} + r^c_{mn}r^b_{nm;a})
\end{equation}
contains all the information of the quantum geometry of the wave functions belonging to bands $n$ and $m$ and how they contribute to the SC conductivity. The prefactor $q^3$ makes the $\tilde{X}^{a;bc}_{nm}(\bk)$ dimensionless, and removes any scaling effect due to the variation of the twist angle. 

An alternative expression of the SC conductivity has been reported in the literature, when the incoming light is polarized along the $b$ direction \cite{chaudhary2022shift}: 
\begin{equation}\label{eq:alt shift current}
    \sigma^{a;bb} = \frac{\pi e^3}{\hbar ^2} \sum_{m,n} \int \dd^2\bk f_{mn}|r^b_{mn}|^2 S^{a;b}_{mn} \delta(\omega-\omega_{mn}),
\end{equation}
which is equivalent to Eq~\eqref{shift current}. 
A quantity called the shift vector, $S^{a;b}_{mn}$, is introduced. It is given by $S^{a;b}_{mn} = r^{a}_{nn} - r^{a}_{mm} - \partial_{k_a} \mathrm{Arg}(r^b_{mn})$. 
Eqs.~\eqref{shift current} and \eqref{eq:alt shift current} are related by the fact that $|r^b_{mn}|^2 S^{a;b}_{mn} = \Im\left[ r^b_{nm} r^b_{mn;a} \right]$. The shift vector is often analyzed due to its gauge-invariance.
In our work, we will show instead the analysis of the gauge-invariant quantity, $\tilde{X}^{a;bc}_{nm}(\bk)$, hereby called ``quantum geometry measure'' throughout this paper, as it is more adapted to Eq~\eqref{shift current} and is more directly related to the amplitude of the SC response.

Although the density of states and the twist angle set a very tempting upper bound of the SC response, we still need to examine another vital element, the quantum geometry, to determine the final SC output. 
To briefly summarize the analysis so far, we have identified three major factors that determine the amplitude of the SC in twisted multilayer systems: Band structure (energy gaps), the quantum geometry and the twist angle. The first two factors are commonly identified in all physical systems having SC response. In addition, we also consider the twist angle as an unusual amplification factor that is decisive to the extraordinary SC conductivity in twistronic systems.

\subsection{Model Hamiltonian}\label{sec model}
In this section we explain the model employed for describing hTTG in different stacking configurations. 
Different from the mirror symmetrical twisted trilayer graphene (mTTG), where the top and bottom layers are perfectly aligned~\cite{caluga2021tstg}, the symmetry and topology of hTTG varies dramatically with different stacking of layers. In fact, mTTG is equivalent to an effective TBG with a decoupled Dirac cone, hence it can only be regarded as an extremely particular case of the TTG system and it does not reflect its fundamental properties such as the supermoir\'e effects and layer-shift induced topological phase transition~\cite{park2021tunable}. As a case study and without losing generality, we limit our discussion to hTTG with equal twist angles, i.e., the top and the bottom layer are rotated to opposite directions with respect to the middle layer, but by the same angle.

As has been pointed out in previous studies\cite{mao2023supermoire, guerci2024chern, guerci2024nature}, the layer stacking in hTTG is crucial for determining the physical properties of the system. Therefore, two phase factors $\phi_1$ and $\phi_2$ related to layer stacking must appear explicitly in the continuum model of hTTG~\cite{trithep2023magic, guerci2024chern}. Assuming the middle layer is unrotated, while the top and the bottom layers are rotated anticlockwise and clockwise, respectively, by the same angle $\theta$, the hTTG Hamiltonian reads
\begin{equation}\label{eq: ham httg}
\begin{split}
&H_K(\br; \phi_1, \phi_2) = m \sigma_z \otimes \mathbb{I}_{\mathrm{layer}} + \\
&\begin{bmatrix}
    \hbar v_F \hat{\bk} \cdot \bm \sigma & w_1 V_{\phi_1,\phi_2}(\br) & \varnothing\\
    w_1 V^\dag_{\phi_1,\phi_2}(\br)  & \hbar v_F \hat{\bk} \cdot \bm \sigma & w_1 V_{-\phi_1,-\phi_2}(\br) \\
    \varnothing & w_1 V^\dag_{-\phi_1,-\phi_2}(\br) & \hbar v_F \hat{\bk} \cdot \bm \sigma 
\end{bmatrix}
\end{split}
\end{equation}
where $v_F$ is the Fermi velocity of single layer graphene, and $\hat{\bm k} = -i \bm \nabla$.  $\bm \sigma = [\sigma_x, \sigma_y]^T$, where $\sigma_{x/y/z}$ are the Pauli matrices defined on the sublattice degree of freedom. $w_1 = 110~\mathrm{meV}$ is the interlayer hopping energy between \emph{different} sublattices. The dimensionless interlayer potential is given by $V_{\phi_1,\phi_2}(\br) = T_1 e^{-i\bq_1 \cdot \br} + T_2 e^{i\phi_1} e^{-i \bq_2 \cdot \br} + T_3 e^{i\phi_2} e^{-i\bq_3 \cdot \br} $. The instance $\phi_1 = \phi_2 = 0$ corresponds to the AAA stacking, while for ABA stacking it is necessary to adopt $\phi_1 = -\phi_2 = 2\pi/3$~\cite{mao2023supermoire, guerci2024chern}.
The $T_j$ matrices describe the interlayer hopping energy between sublattices, and are defined by
\begin{equation}\label{eq: Tj matrix}
    T_j = \begin{bmatrix}
        r & e^{-2i (j-1)\pi/3} \\
        e^{2i(j-1)\pi/3} & r
    \end{bmatrix} ,
\end{equation}
with the real number $r \in [0,1]$ usually called the corrugation parameter. $r w_1$ designates the interlayer hopping amplitude between the \emph{same} sublattices. $r=0$ is tagged as the \emph{chiral} limit of twisted bilayer graphene, while $r=1$ signifies the \emph{isotropic} limit. 
These Hamiltonians follow from the reasoning of the model featuring TBG, which the reader can find in Appendix \ref{app:tbgmodel} for more details.
In addition, Eq.(\ref{eq: ham httg}) has inversion symmetry breaking inherited from the tight-binding model of monolayer graphene with hBN. An additional sublattice offset $m \sigma_z \otimes \mathbb{I}_\mathrm{layer}$ is added to the Hamiltonian to mimic the influence of the substrate on which the trilayer graphene is deposited. $m$ is the offset energy, while $\mathbb{I}_\mathrm{layer}$ is the identify matrix on the \emph{layer} indices. 
The sublattice offset opens the gap at the Dirac point connecting the middle bands, but preserves the $C_{3z}$ symmetry of the system. However, it breaks the particle-hole symmetry (PHS), $C_{2x}$ and $C_{2z}T$ symmetries.

\section{Results and Discussion}\label{sec results}

In this section, we provide the numerical results of the SC conductivity in hTTG. We first focus on the equally twisted ABA-stacked hTTG, with the corrugation parameter set to $r = 0.8$, which is closest to the real-life scenario~\cite{koshino2018}. Detailed analysis of the SC response in this setup is performed at the magic angle ($1.95^\circ$). For the sake of comparison, we present later the results of different twist angles and different stacking. The chiral limit ($r = 0$) is a theoretically interesting scenario and is also investigated. 
As a way to complement these results, we share the outcomes for the SC conductivity in gapped single layer graphene and in TBG in Appendices~\ref{sc in graphene} and ~\ref{app:matbg}, respectively.

\subsection{Shift current in ABA stacked magic-angle helical twisted trilayer graphene}\label{sec: sc aba httg}

As the ABA stacked trilayer graphene is believed to be energetically more favorable~\cite{trithep2023magic}, we develop a detailed in-depth investigation to this configuration at the magic angle.

\begin{figure}[!htbp]
    \centering
    \includegraphics[width=\linewidth]{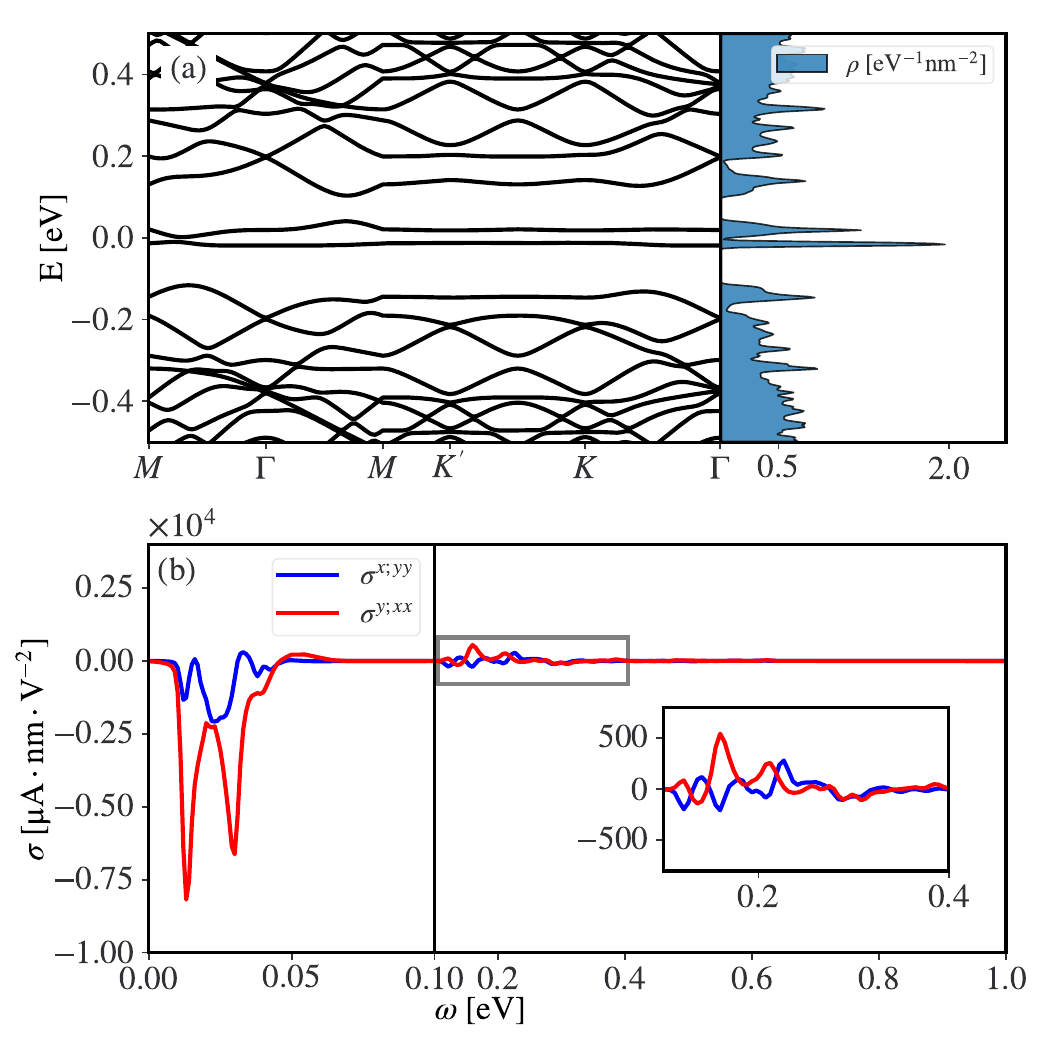}
    \caption{
    (a) Band structure and the associated density of states of an equal-twist ABA-hTTG at the first magic angle ($1.95^\circ$), where the two middle bands are not completely flat due to the deviation from the chiral limit.
    (b) The shift-current conductivity computed with $T = 0$ and $\mu = 0$ for $\sigma^{x;yy}$ (blue lines) and $\sigma^{y;xx}$ (red lines) as functions of the incident photon energy. The plot zooms in the THz photon energy range below $0.1$ eV. The inset plot zooms in the conductivity plot within the gray rectangle. 
    }
    \label{fig:aba band dos sc}
\end{figure}

Firstly, for the case of hTTG at first magic angle $1.95^\circ$, we illustrate the band structure, the density of states and the conductivity coefficients $\sigma^{x;yy}$ and $\sigma^{y;xx}$ in Fig.~\ref{fig:aba band dos sc}. We remark that the model of hTTG is constrained by C$_{3z}$, where $(x,y)\to \frac{1}{2}(-x-\sqrt{3}y,\sqrt{3}x-y)$. Then, the conductivity coefficients for this system comprises two independent components $\sigma^{x;yy}$ and $\sigma^{y;xx}$ satisfying the relations \cite{Chen2024}
\begin{equation}
\begin{split}
\sigma^{x;yy}=-\sigma^{x;xx}=\sigma^{y;yx}=\sigma^{y;xy}\not=0,\\
\sigma^{y;xx}=-\sigma^{y;yy}=\sigma^{y;xy}=\sigma^{x;yx}\not=0.
\end{split}
\end{equation}

Regarding Fig.~\ref{fig:aba band dos sc}(a), we remark that in hTTG completely flat bands are possible only at the chiral limit $r=0$. In the realistic case with $r = 0.8$, the magic angle is defined by zero-dispersion velocity at the $\mathbf{K}$ or $\mathbf{K}'$ points of the moir\'e Brillouin zone, while the middle bands near the Fermi level cannot be entirely flattened. Notice that varying the corrugation parameter $r$ also leads to non-negligible variation of the magic angle, as explained in Appendix~\ref{app:magic angle}. 
If we look at the two bands closest to the Fermi level and according to Fig.~\ref{fig:aba band dos sc}(a), there is a large density of states at the middle flattened bands corresponding to the charge neutrality point. 
As we have set the chemical potential to be exactly at the zero-energy level, one of the middle bands is completely filled while the other is entirely empty. As we can get from Fig.~\ref{fig:aba band dos sc}(b), the transition between the two middle bands leads to an enormous SC conductivity of about $\sim 10^4~\mathrm{\mu A \cdot nm \cdot V^{-2}}$. The strong signal found in the terahertz (THz) regime could allow for the experimental measurement of the band geometry and possibly topology with high signal-noise ratio~\cite{kumar2024terahertz}. 
The density of states exhibited in Fig.~\ref{fig:aba band dos sc}(a) also features consecutive peaks in the density of states that coincide with the regions away from the middle bands. 
Therefore, the transition between the middle bands and these higher-energy bands gives rise to significant SC response to photons with energy from $0.1$ to $0.3~\mathrm{eV}$, corresponding to the mid-infrared light frequency. 
Although this photoconductivity is of order of $\sim 10^2~\mathrm{\mu A \cdot nm \cdot V^{-2}}$, it is still considerable as a second-order response to the electric field.
The significant SC response to infrared light in hTTG is in great contrast to TBG \cite{chaudhary2022shift, kumar2024terahertz}, whose SC response to mid-infrared light is almost negligible (See Appendix \ref{app:matbg} for more details).

\begin{figure}[!htbp]
    \centering
    \includegraphics[width=0.95\linewidth]{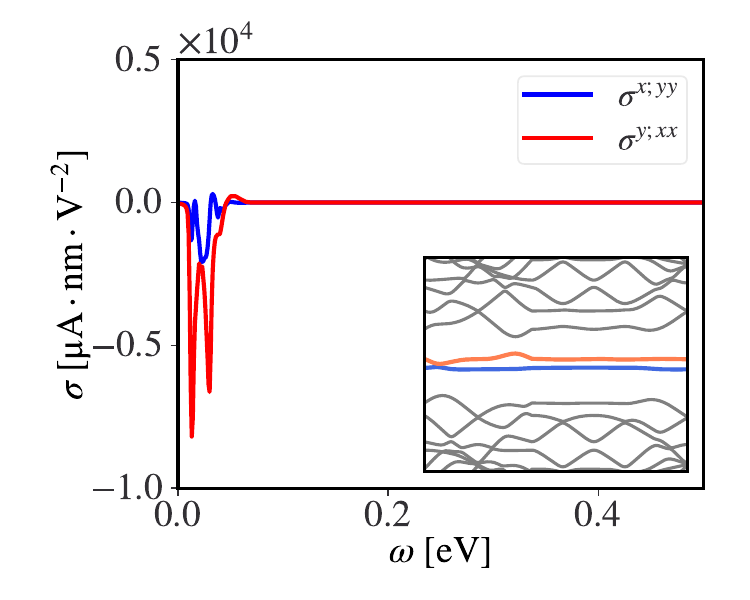}
    \caption{ SC conductivity from transitions between flat bands, with $\sigma^{x;yy}(\omega)$ and $\sigma^{y;xx}(\omega)$ colored in blue and red lines, respectively. The inset presents schematically the band structure, highlighting the two bands involved in the main plot, with the occupied band colored in light blue and the unoccupied band in orange.}
    \label{fig:sc flat to flat}
\end{figure}

In what follows, since Eq.~\eqref{shift current} is a sum over band pairs $(n,m)$, let us examine the interband contributions to the SC arising from transitions between different bands. Our results are reported in Fig.~\ref{fig:sc flat to flat}, where the main plot depicts the SC conductivity due to the transition between the two middle flat bands. In comparison with Fig.~\ref{fig:aba band dos sc}, it is clear that the SC response within the far-infrared (THz) regime is entirely accounted for by the transition between the two close middle bands. The high density of states and small separation in energy greatly promote the THz SC response. In contrast, this transition does not contribute to the SC conductivity for $\hbar\omega \geqslant 0.1$ eV.

\begin{figure}[!htbp]
    \centering
    \includegraphics[width=\linewidth]{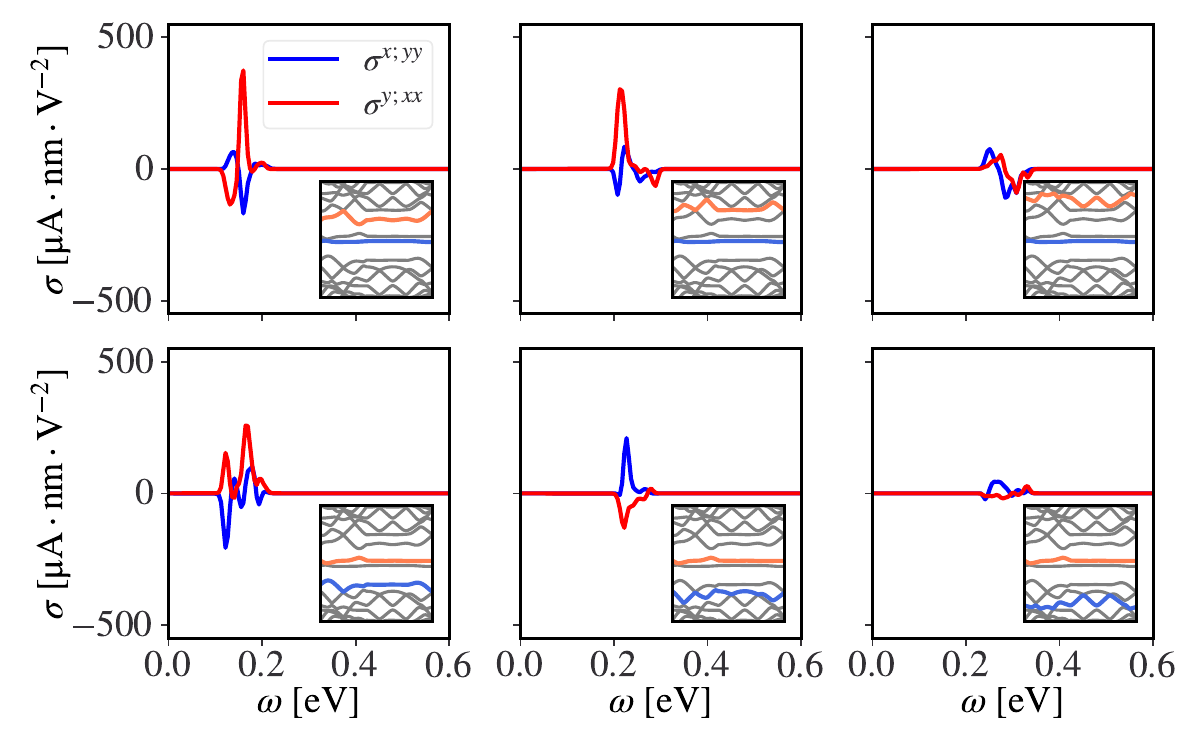}
    \caption{
    SC conductivity coefficients $\sigma^{x;yy}$ (blue line) and $\sigma^{y;xx}$ (red line) from transitions between flat bands and dispersive bands. The first row shows the SC contribution by transition from the occupied flat band to unoccupied dispersive bands, while the the SC conductivity presented in the second row is from the transition from occupied dispersive bands to the unoccupied flat bands. The insets show schematically the band structure while highlighting the bands involved in each main plot. The occupied and the unoccupied bands are colored in light blue and orange, respectively.
    }
    \label{fig:sc aba flat to disp}
\end{figure}

In addition, in Fig.~\ref{fig:sc aba flat to disp} we demonstrate how the transitions between flattened bands and dispersive bands contribute to the SC response. Combining Figs.~\ref{fig:sc flat to flat} and~\ref{fig:sc aba flat to disp}, we notice that they are mainly responsible for the SC signals in the mid-infrared regime, but they do not contribute to the far-infrared THz range. 
We have also computed the SC conductivity from the transition between dispersive bands, finding smaller values of the SC conductivity of orders of $10~\mathrm{\mu A \cdot nm \cdot V^{-2}}$. Therefore, we confirm that the flattened middle bands in hTTG are the key ingredient to obtain the extraordinary SC response of this assembly in both far and mid infrared regimes. The origin of this signal is associated with the high density of states due to the flattened bands near the Fermi level. Moreover, the SC conductivity in the THz regime is further augmented by the small energy separation between the two middle bands. 

We notice that the far-infrared THz SC response is about 2 orders of magnitude greater than the SC response to mid- and near-infrared light. Such a substantial difference cannot be accounted for by these two simple reasons. As pointed out at the end of Section~\ref{sec: sc amplification}, the internal structure of the wave functions has to play a role to magnify the SC response by two orders of magnitude. The only ingredient left to be examined is the dimensionless quantity serving as a measure of the quantum geometry defined in Eq.~\eqref{eq: dimless measure}. 

\begin{figure}[!htbp]
    \centering
    \includegraphics[width=\linewidth]{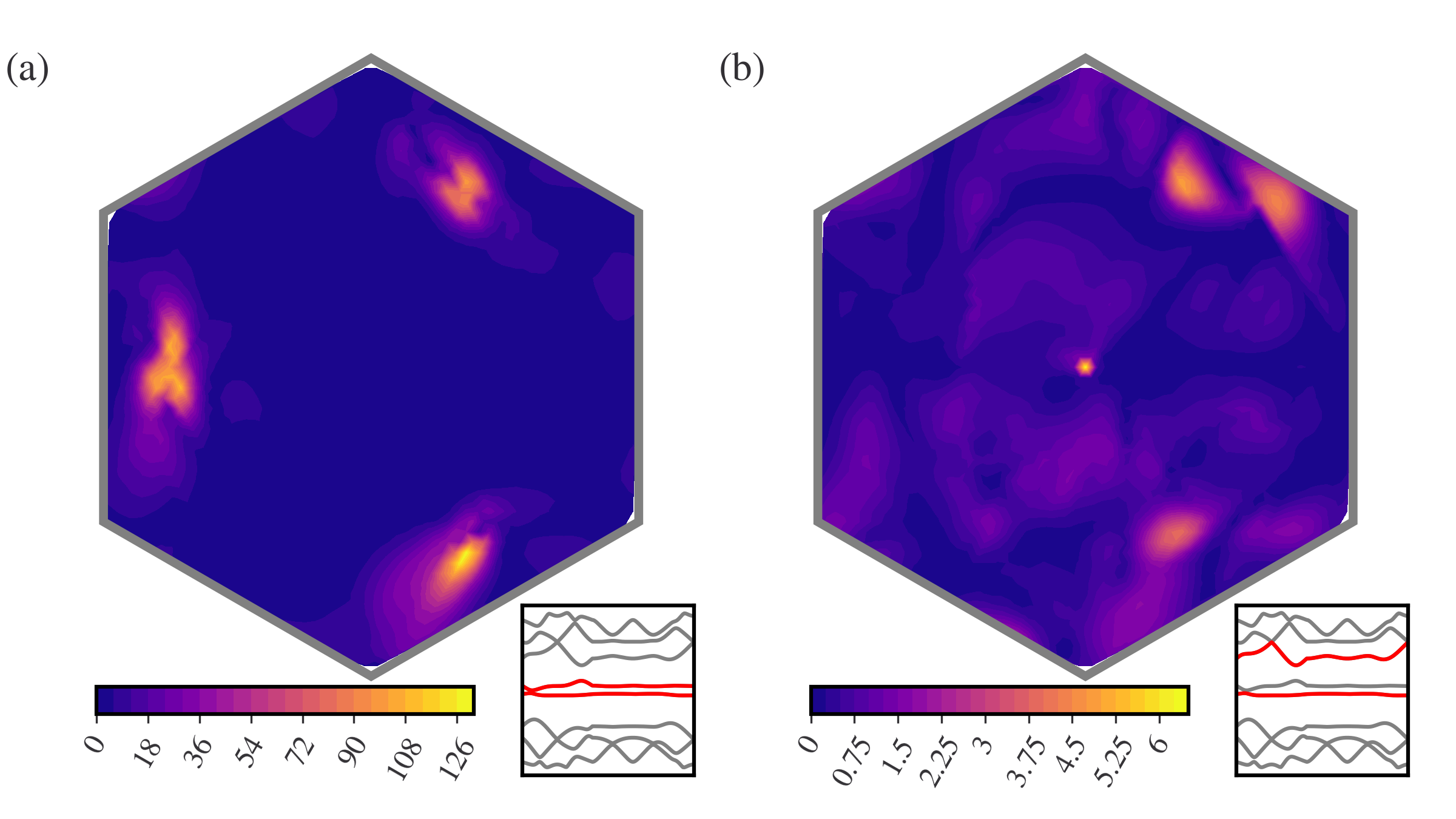}
    \caption{ 
    Dimensionless quantum geometry measure $\tilde{X}^{x;yy}(\bk)$ within the moir\'e Brillouin zone computed for (a) the two middle flattened bands and (b) a middle flat band to dispersive band. Both color plots uses different color scales as indicated by the colorbar beneath each subfigure. The bands contributing to the quantum geometry measure are highlighted with red color in each inset. In both color plots, the moir\'e Brillouin Zone is marked with gray lines.
    }
    \label{fig:geometry}
\end{figure}

Indeed, the quantum geometry measure between the two middle bands promotes the THz SC by another order of magnitude with respect to the mid-/near-infrared SC.

Before our work, S. Chaudhary et al. have reported exceptional SC response in magic-angle twisted \emph{bilayer} graphene (MATBG)~\cite{chaudhary2022shift}. Comparing the SC response found in hTTG with their data without electron interactions, the THz SC response in hTTG is even $1$ to $2$ orders of magnitude higher than MATBG. 
In order to make a comparative analysis of the two systems, we also compute the SC response of MATBG with the same corrugation factor $r=0.8$, at zero-temperature and with the chemical potential set to $\mu = 0$. The gap opening sublattice offset potential is also set to $m=30~\mathrm{meV}$, same as in hTTG.
The model details of TBG of this work are explained in Appendix~\ref{app:tbgmodel} and the results are presented in Appendix~\ref{app:matbg}. 
Comparing Fig.~\ref{fig:aba band dos sc} and Fig.~\ref{fig:matbg sc normal}, the differences in band gap and the density of states cannot fully explain the notorious difference in the orders of magnitude of SC response. The only explanation comes from the quantum geometry measure. 
Indeed, comparing Fig.~\ref{fig:geometry}(a) and Fig~\ref{fig:tbg quantum geometry}(a), the quantum geometry measure of the middle bands in hTTG is almost 2 orders of magnitude greater than in TBG. 

We also remark that in magic-angle hTTG, the energy gap, density of states and the quantum geometry measure are all in favor of high SC response in the THz regime. In contrast, the quantum geometry measure of the two middle bands in magic angle TBG is in fact weaker than that between the middle band and a dispersive band. The exceptional SC response in MATBG is rather supported by the small gap opening and the high density of states in the middle flattened bands.

\subsection{Shift current in ABA-stacked hTTG away from the magic angle}

In the following, we explore the scenarios where the system is tuned away from the magic angle and the stacking of graphene layers is altered. We show that the large SC response is not limited to magic angles, but rather to small angles. 

\begin{figure}[!htbp]
    \centering
    \includegraphics[width=\linewidth]{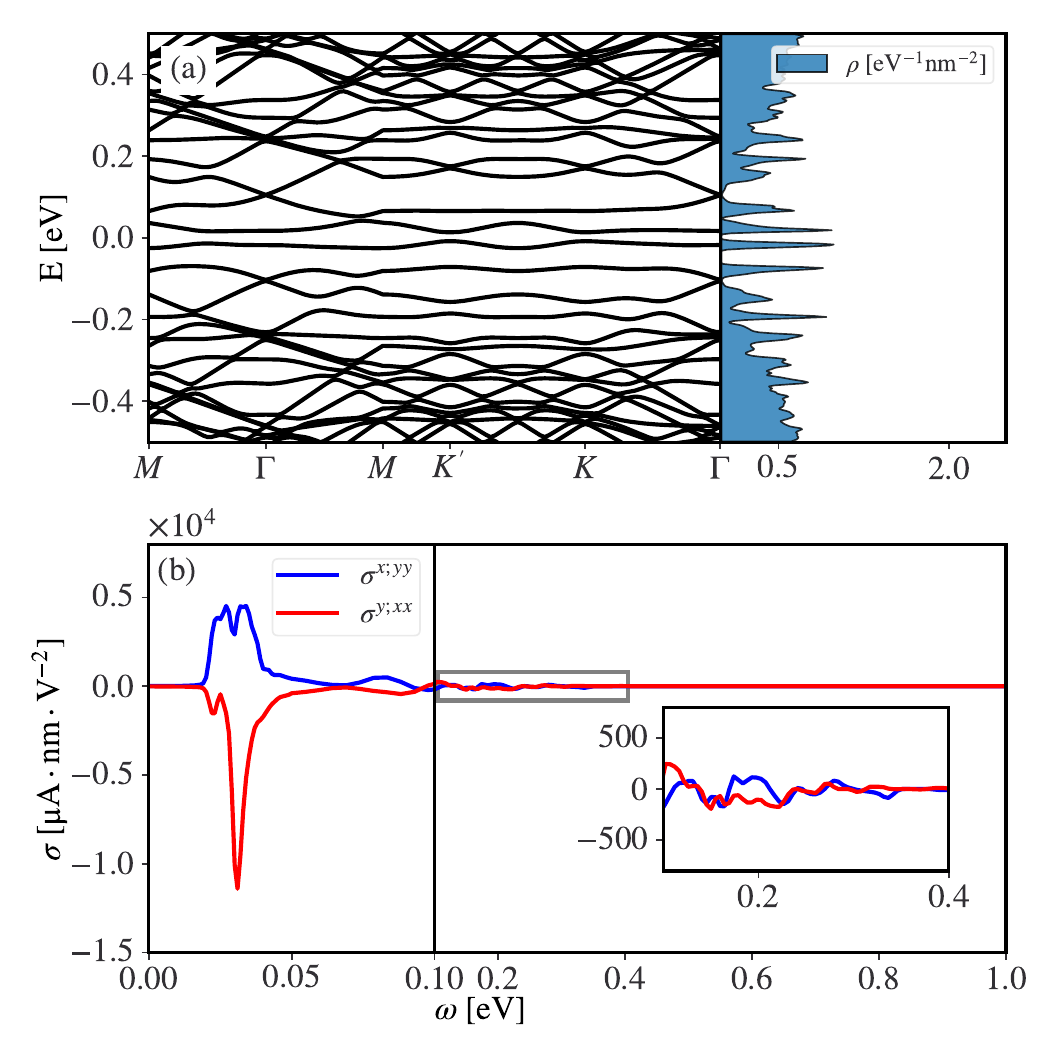}
    \caption{
    (a) Band structure and the associated density of states of an equal-twist ABA-hTTG with equal twist angles being $1.5^\circ$.
    (b) The shift-current conductivity computed with $T = 0$ and $\mu = 0$ for $\sigma^{x;yy}$ (blue lines) and $\sigma^{y;xx}$ (red lines) as functions of the incident photon energy. The plot zooms in the THz photon energy range below $0.1$ eV. The inset plot zooms in the conductivity plot within the gray rectangle.
    }
    \label{fig:aba sc 1.5deg}
\end{figure}

\begin{figure}[!htbp]
    \centering
    \includegraphics[width=\linewidth]{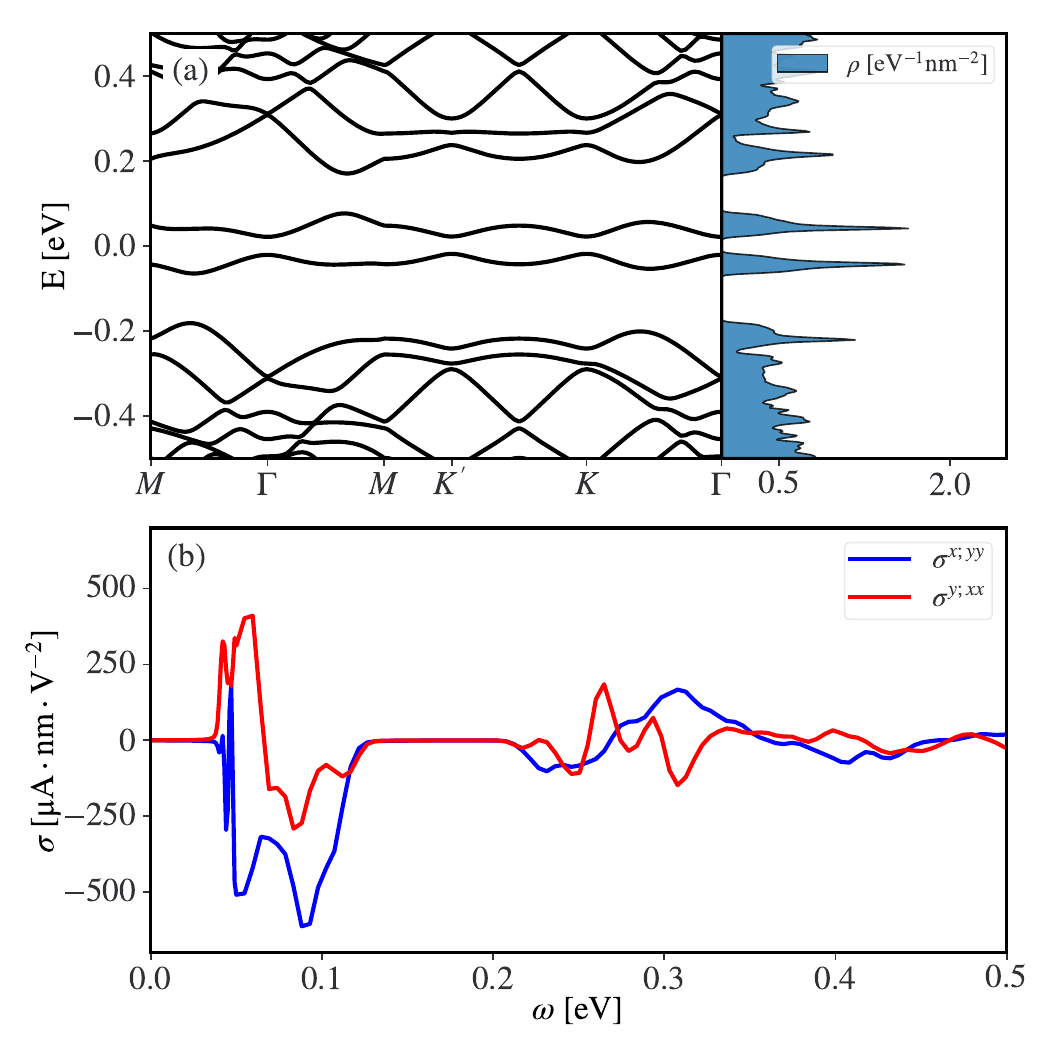}
    \caption{
    (a) Band structure and the associated density of states of an equal-twist ABA-hTTG with twist angles being $2.5^\circ$.
    (b) The shift-current conductivity computed with $T = 0$ and $\mu = 0$ for $\sigma^{x;yy}$ (blue lines) and $\sigma^{y;xx}$ (red lines) as functions of the incident photon energy.} 
    \label{fig:aba sc 2.5deg}
\end{figure}

Tuning the twist angle away from the magic angle, the dispersion velocity near the $\mathbf K$ points is restored, and it is expected to reduce the density of states related to the flattened bands. We consider two directions of deviation from the magic angle: increasing and reducing it, respectively. Then, we illustrate the SC conductivity coefficients for the twist angles $1.5^\circ$ in Fig.~\ref{fig:aba sc 1.5deg} and $2.5^\circ$ in Fig.\ref{fig:aba sc 2.5deg}. 
The revival of the dispersion velocity in the middle bands leads to significant reduction of density of states near the Fermi level in both cases. 

Notice that the SC conductivity profiles described in Figs.~\ref{fig:aba sc 1.5deg} and \ref{fig:aba sc 2.5deg} reveal a remarkable contrast between both instances.
On one side, Fig.~\ref{fig:aba sc 1.5deg} suggest that reducing the twist angle to $1.5^\circ$ does not have a dramatic impact on the extraordinary SC conductivity observed in the far-infrared regime. Besides, the signal identified in the mid-infrared region for the first magic angle has been shifted to THz frequencies due to stronger band compression with smaller twist angle. In spite of the reduced density of states in the middle bands, lowering the twist angle causes stronger compression of the whole band structure to small energy scales, as signified by Eq.~\eqref{eq:ham tbg dimless}. As pointed out by Eqs.~\eqref{eq: connection} and \eqref{eq: gderiv connection}, the interband connections and their generalized derivatives also gain substantial amplification due to smaller band separations $\omega_{nm}$. 
Therefore, the influence of the reduced density of states on the SC conductivity is completely compensated by the band compression.
On the other hand, in Fig.~\ref{fig:aba sc 2.5deg} we can see that with a twist angle of $2.5^\circ$ the SC conductivity is weakened by both the reduced density of states and the widened band gap simultaneously. This leads to a decrease in the SC conductivity at THz frequencies to an order of $100~\mathrm{\mu A \cdot nm \cdot V^{-2}}$ [Fig.~\ref{fig:aba sc 2.5deg}(b)]. In this case, the mid-infrared SC response shifts towards the near-infrared region due to the band decompression to higher energy scales with large twist angles.

\subsection{Chiral suppression of shift current}
\begin{figure}[!htbp]
    \centering
    \includegraphics[width=\linewidth]{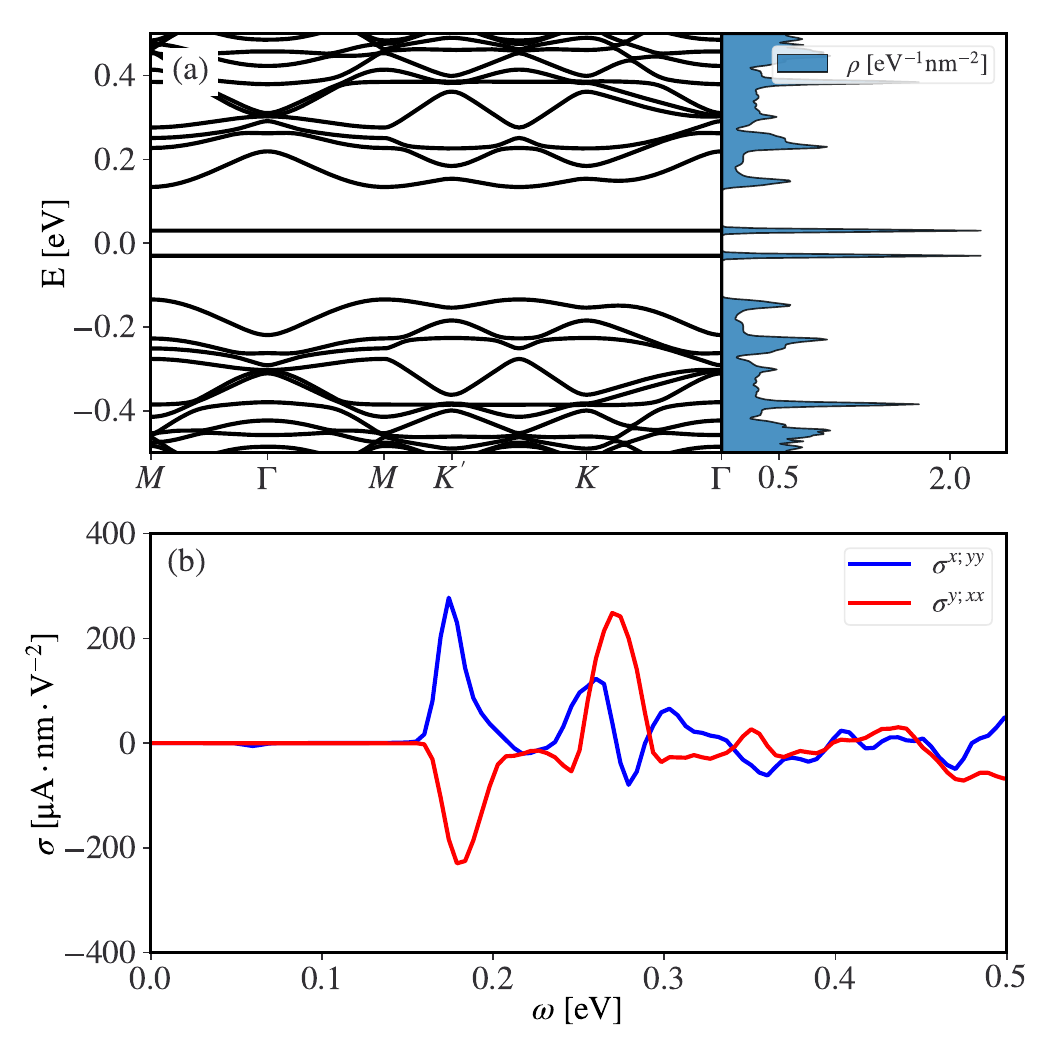}
    \caption{
    (a) Band structure of the magic-angle equal-twist ABA-hTTG at \emph{chiral} limit with the associated density of states, where the chiral magic angle is $1.69^\circ$; (b) Shift current conductivity tensors $\sigma^{x;yy}$ (blue curve) and $\sigma^{y;xx}$ (red curve) as functions of the incident photon energy.
    }
    \label{fig:aba sc chiral}
\end{figure}

The chiral limit of twisted bilayer and trilayer graphene is a useful and convenient scenario for the theoretical study of these structures, as it allows for the retrieval of analytical flat-band wave functions~\cite{tarnopolsky2019origin, guerci2024chern, guerci2024nature}. Nevertheless, in this section we can show that the SC response with chiral flat-band wave functions is obstructed by their sublattice polarization. In Fig.~\ref{fig:aba sc chiral} we plot the band structure, density of states and SC conductivity profiles of hTTG with ABA stacking at the chiral limit, where the magic angle is $1.69^\circ$. 

As can be deduced from Fig.~\ref{fig:aba sc chiral}(b), the THz SC response due to the transition between the two flat bands is almost depleted, despite the extraordinary high density of states in the perfectly flat bands and the compression of the band structure, which are both in favor of high SC response in the THz regime. We can explain the suppression of the THz SC response from the fact that the flat-band wave functions are sublattice-polarized, i.e., they only have non-zero components on one of the sublattices of graphene. As a direct consequence, the interband connection between the chiral flat bands $r^a_{nm} = i\mel{u_n}{\partial_{k_a}}{u_m}$ is \emph{strictly} zero, with $n, m$ denoting the band indices of the flat bands. Notice that the cancellation of the interband connection between middle bands is gauge-independent and cannot be saved by gauge change. Besides, the wave functions from dispersive bands are not sublattice polarized, thus the interband connection between a flat band and a dispersive band survives at the chiral limit. However, to the best of our knowledge, no analytical expression is available for wave functions of other bands than the flat bands~\cite{tarnopolsky2019origin, guerci2024chern}.
Our results reveal that the handy analytical tools of chiral wave functions~\cite{tarnopolsky2019origin, guerci2024chern, guerci2024nature} are not appropriate for the investigation of the SC response. The chiral suppression of SC conductivity is a direct evidence that the quantum geometry and the structure of the wave functions play an essential role in the nonlinear optical response. The results from the chiral hTTG suggest that sublattice polarization is a strong impeding factor of SC generation and should be ruled out in future search of high SC response in quantum materials.

\subsection{Shift current in magic-angle AAA stacked helical twisted trilayer graphene}\label{sec: sc aaa httg}

\begin{figure}[!htbp]
    \centering
    \includegraphics[width=\linewidth]{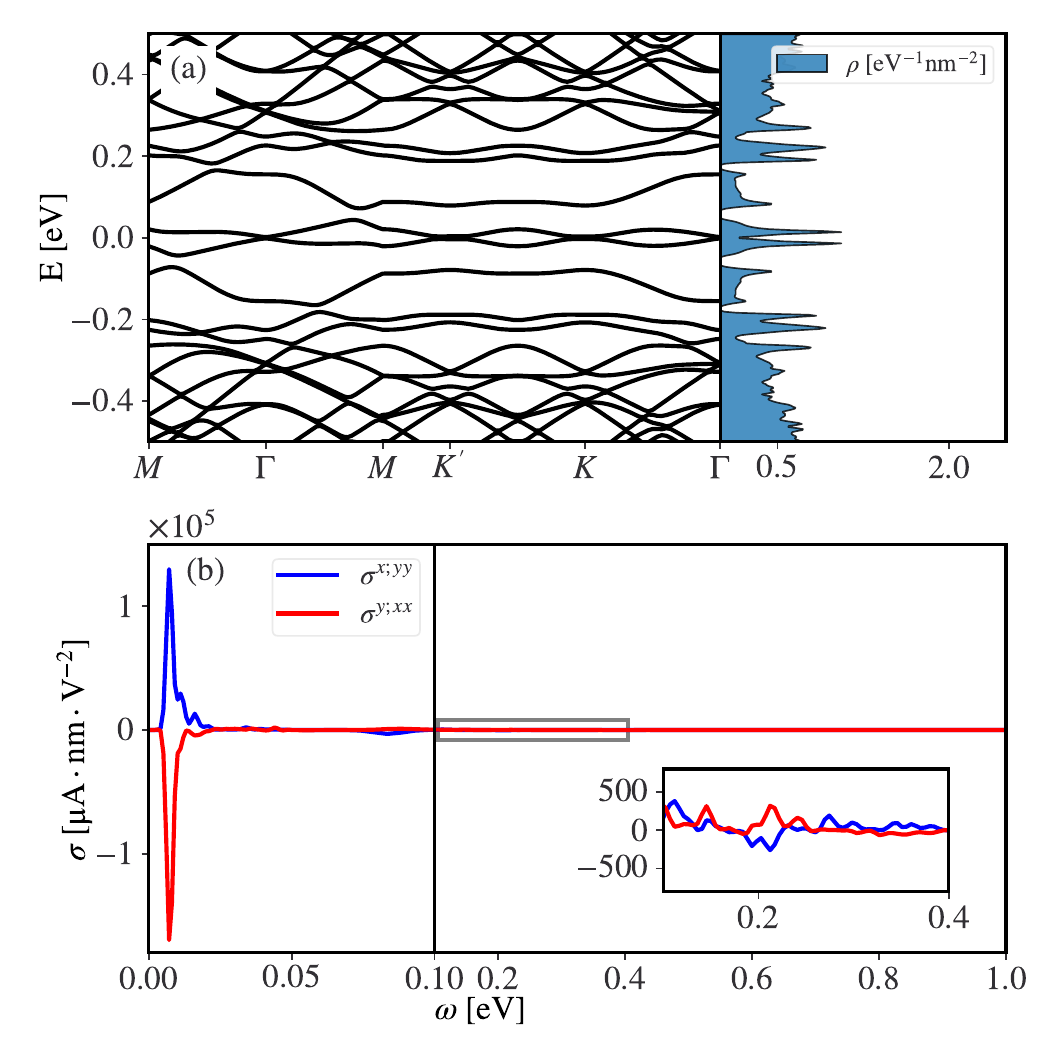}
    \caption{
    (a) Band structure of an equal-twist AAA-hTTG at first magic angle ($1.75^\circ$), and the associated density of states within the same energy range as in the band structure plot. 
    (b) The shift-current conductivity computed with $T = 0$ and $\mu = 0$ for $\sigma^{x;yy}$ (blue lines) and $\sigma^{y;xx}$ (red lines) as a function of the incident photon energy. The plot zooms in the THz photon energy range below $0.1$ eV. The inset plot zooms in the conductivity plot labeled with the gray rectangle.}
    \label{fig:aaa band dos sc}
\end{figure}

The tunability via layer stacking is an advantage of the trilayer graphene system in contrast to the twisted bilayer graphene. In this section, to show how the SC response is altered by different layer stacking, we consider the AAA-stacked hTTG and we compare it with the ABA stacked hTTG in previous sections. We report the band structure, the density of states and the SC conductivity coefficients in Fig.~\ref{fig:aaa band dos sc}.

In comparison to the ABA stacked case, the band structure displayed in Fig.~\ref{fig:aaa band dos sc}(a) has a larger number of crossings between the two closest bands to the Fermi level. Moreover, according to Fig.~\ref{fig:aaa band dos sc}(a), the profile of the density of states is similar to the case mentioned in Fig.~\ref{fig:aba sc 1.5deg}(a). Remarkably, whereas Fig.~\ref{fig:aaa band dos sc}(b) displays a SC response at mid-infrared regime of order $10^2~\mathrm{\mu A \cdot nm \cdot V^{-2}}$, the SC photoconductivity in the THz regime has a maximum to the order of $10^5~\mathrm{\mu A \cdot nm \cdot V^{-2}}$. In principle and to the best of our knowledge, this outcome suggests that AAA stacked hTTG could be a platform of an unprecedented SC conductivity, surpassing the values documented in the literature.

\subsection{Shift current at finite temperature}
Our discussion on the SC response so far has been focusing on zero temperature. To complete our discussion and to show the performance of the system at finite temperature, we extend our calculations to different temperatures, for the ABA stacked hTTG with equal twist angle being $1.95^\circ$. 

\begin{figure}[!htbp]
    \centering
    \includegraphics[width=\linewidth]{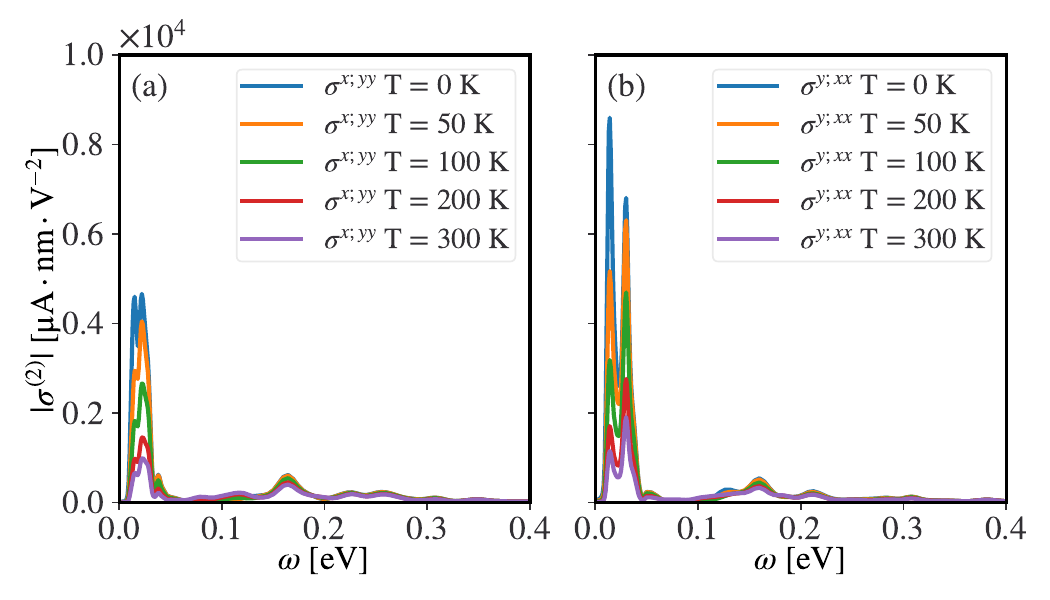}
    \caption{
    The shift current conductivity of ABA hTTG at different temperature. The twist angles are fixed at $1.95^\circ$, and the chemical potential is set to be $0$. To ease the comparison, the absolution values of shift current conductivity are shown. $\sigma^{x;yy}$ and $\sigma^{y;xx}$ are exhibited in (a) and (b), respectively.
    }
    \label{fig:sc vs T}
\end{figure}

In Fig.~\ref{fig:sc vs T}, we show the amplitude of the SC conductivity $\sigma^{x;yy}$ and $\sigma^{y;xx}$ calculated at different temperature, ranging from $0$ K to $300$ K.  Up to room temperature, only the population of electrons in the two middles bands are significantly altered. The higher bands are separated from the middle bands by an energy gap $\sim 0.15$ eV, which corresponds to a temperature level of $\sim 1700$ K. Therefore, only the SC response related to the transition between the middle bands is substantially affected by raising the temperature.
Increasing the temperature augments the population of electrons in the conduction band of the middle bands, while the occupation number in the valence band of the middle bands is reduced. We know from Eq.~\eqref{shift current} that the \emph{difference} of the occupation number between the bands determines the amplitude of the SC response. The THz SC response is consequently dampened due to the reallocation of electrons in the two middle bands when the temperature increases. However, the THz SC response stays at an order as high as $10^3~\mathrm{\mu A \cdot nm \cdot V^{-2}}$, showing that this material retains significant SC performance even at room temperature. 

\section{Conclusion}\label{conclusion}

To summarize, we identify hTTG as an ideal platform for the observation of a highly controllable SC conductivity in the infrared frequency regime. In this context, we deduce that the twist angle rather than only the magic angle, is essential for a gigantic SC conductivity in the THz regime. Moreover, the compression of the band structure and the flattening of the middle bands enhance the density of states near the Fermi level ($\mu=0$),further promotes the far-infrared THz SC response up to orders of $10^4~\mathrm{\mu A \cdot nm \cdot V^{-2}}$ and robust mid-infrared SC response in orders of $10^2~\mathrm{\mu A \cdot nm \cdot V^{-2}}$. 
We conclude that the large density of states arising from the flattened middle bands is crucial to the magnification of SC, but the final output is limited by the quantum geometry of wave functions. Moreover, altering layer stacking also impacts dramatically the magnitude of the SC conductivity. All these findings encourage further theoretical and experimental studies on hTTG as a realistic scenario for SC generation with high efficiency.

\section*{Acknowledgments}

C.A. and Y.M. acknowledge ANR project COLIBRI No. ANR-22-CE30-0027. D.G.O. thanks the support of the Excellence Initiative of Aix-Marseille Universit\'e - A*Midex, a French ``Investissements d'Avenir'' program.

\appendix

\section{Shift current in gapped graphene single layer}\label{sc in graphene}
\begin{figure}[!htbp]
    \centering
    \includegraphics[width=\linewidth]{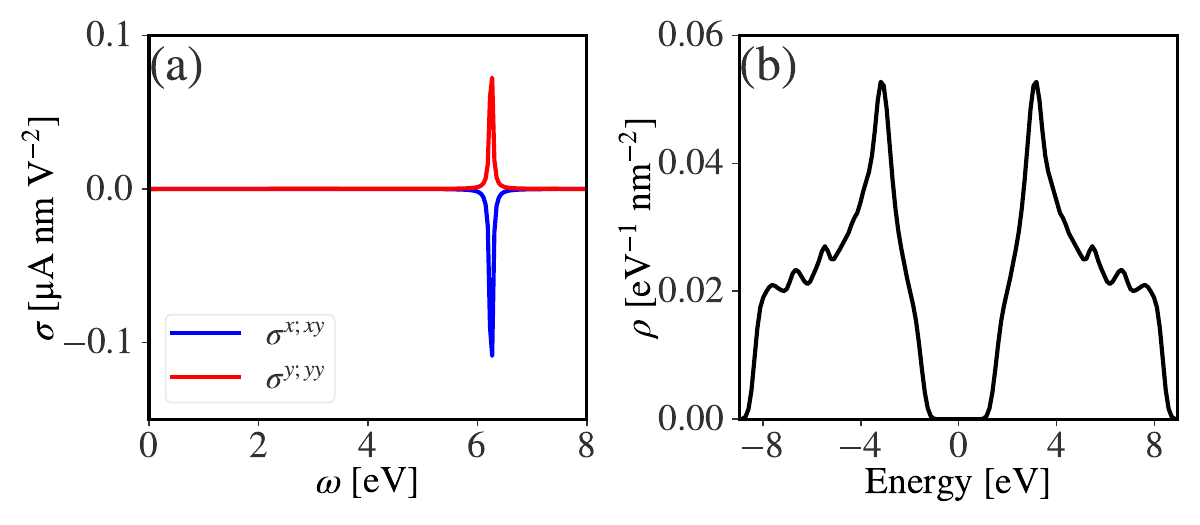}
    \caption{(a) SC conductivity coefficients of the gapped SLG, and (b) density of states. The location of the peak of the SC conductivity corresponds exactly to the distance of the highest peaks in density of states. The peaks in DOS profile originate from the flattened parts of the band structure near the $\mathbf{M}$-points of graphene.}
    \label{fig:sc dos graphene}
\end{figure}

To show that the SC conductivity can be bounded by the high density of state peaks in the conduction band and the valence band, the SC conductivity coefficients are computed with a toy model based on a gapped single-layer graphene (SLG). The tight-binding model employed for the gapped SLG is given by \cite{CastroNeto2009}
\begin{equation}\label{graphene}
\begin{split}
    \hat H_\text{SLG} = \sum_{\bR} \sum^3_{j = 1} -t c^\dag_{\bR, A} c_{\bR+\bm d_j, B} + \mathrm{h.c.} \\
    + \sum_{\bR} m (c^\dag_{\bR,A} c_{\bR,A} - c^\dag_{\bR+\bm d_1, B} c_{\bR+\bm d_1, B}) ~,
\end{split}
\end{equation}
where $\bR$ is a vector of the triangular Bravais lattice that coincides with the carbon atoms belonging to the sublattice $A$, and $d_j = d [-\sin(2(j-1)\pi/3), \cos(2(j-1)\pi/3)]^T$ with $d = 1.42$ \AA being the carbon-carbon bond length. We have adopted $t = 2.73~\mathrm{eV}$. The mass $m$ appearing in the second term of Eq.~\eqref{graphene} is essential to break the time reversal symmetry so as to have significant SC response. In this toy model we have employed $m = 1.5~\mathrm{eV}$ to widen the gap. 
Only the coupling between nearest neighbors is taken into consideration. 
It is known that the tight-binding model describes typically a graphene single layer deposited on a hexagonal boron-nitride substrate. Such a Hamiltonian described in Eq.~\eqref{graphene} belongs to the point group $C_{3v}$ and has inversion symmetry breaking. The lack of inversion symmetry is indispensable for a non-vanishing SC conductivity\cite{hipolito2016nonlinear}.
The dominant SC conductivities are located at the light frequency corresponding exactly to interband gap located at $\mathbf{M}$-points of graphene, as shown in Fig.~\ref{fig:sc dos graphene}. The results shown in Fig.~\ref{fig:sc dos graphene} clearly stands for our reasoning in Eq.~\eqref{sc upper lim}.

\section{Twisted bilayer graphene model}\label{app:tbgmodel}

Here we consider a twisted bilayer graphene (TBG) created by the rotation of two graphene layers that are aligned in opposite directions: the top layer is rotated anticlockwise by $\theta/2$ while the bottom layer is rotated clockwise by the same angle. 
The twist angle between the two layers is $\theta$. Then the continuum Hamiltonian of TBG can be developed using the states near the $\mathbf{K}$ or $\mathbf{K}'$ points of each graphene layer\cite{bistrizer2011moire}.
The TBG Hamiltonian developed near the $\mathbf{K}$ points of graphene writes 
\begin{equation}\label{ham tbg}
    H_K(\bm r) = \begin{bmatrix}
        \hbar v_F \hat{\bm k} \cdot \bm \sigma + m \sigma_z & w_1 V(\br) \\
        w_1 V^\dag(\br) & \hbar v_F \hat{\bm k} \cdot \bm \sigma + m \sigma_z
    \end{bmatrix} ,
\end{equation}
The interlayer hopping potential of TBG, $V(\br)$, is given by 
\begin{equation}
    V(\br) = \sum_{j=1}^3 T_j e^{-i \bm q_j \cdot \br},
\end{equation}
The TBG continuum model near the $\mathbf{K}'$ point of graphene can be obtained by applying the time reversal operation on Eq.~\eqref{ham tbg}. In this context,  
$\mathbf{K}$ and $\mathbf{K}'$ of graphene are referred to as the \textit{valley} degree of freedom of the electrons in TBG.

The physics of TBG is uniquely controlled by the twist angle. This can be seen by extracting the typical energy $\hbar v_F q$ in Eq.~\eqref{ham tbg} and substituting the operator $\hat{\bk}$ with a dimensionless operator $\hat{\bk} = - \frac{i}{q}\bm \nabla$, which gives
\begin{equation}\label{eq:ham tbg dimless}
    H_K(\bm r) = \hbar v_F q \begin{bmatrix}
        \hat{\bk} \cdot \bm \sigma + m'\sigma_z & a V(\br) \\
        a V^\dag(\br) &  \hat{\bm k} \cdot \bm \sigma + m'\sigma_z
    \end{bmatrix} ,
\end{equation}
where $a = w_1/(\hbar v_F q) = w_1/(\hbar v_F |\mathbf{K}| \theta)$ and $m' = m/(\hbar v_F q)$. The eigenenergies and eigenstates of TBG is uniquely determined by the dimensionless Hamiltonian $H_K(\br)/(\hbar v_F q)$.  $a  \propto 1/\theta$ is the only parameter that controls the physics of the TBG Hamiltonian. Since the computation of the SC conductivity includes explicit derivative with respect to the wave vector $\bk$ inside the Brillouin Zone, it is convenient to apply the Bloch's theorem to the system and work with the Hamiltonian defined in the reciprocal lattice, parameterized by $\bk$. 
As the $\mathbf{K}'$ model is the time-reversed version of $H_K$, the real parts of $\sigma^{a;bc}$ computed in both valleys will have opposite signs. When electron filling in the system does not distinguish between the valleys in TBG, the SC will be largely canceled out. However, as pointed out in Refs~\cite{sharpe2019emergent,serlin2020intrinsic, ying2021current}, real-life TBG deposited on hBN substrate can be spontaneously valley-polarized. The polarization can be switched conveniently by application of a small DC current. It is therefore meaningful and realistic to consider the SC response from only one of the valleys.  
The SC computed from one valley can be regarded as the largest possible scenario in a real-life TBG system. The actual output will also depend on the valley polarization of the system.

\section{Shift current in magic-angle twisted bilayer graphene}\label{app:matbg}
\begin{figure}[!htbp]
    \centering
    \includegraphics[width=0.95\linewidth]{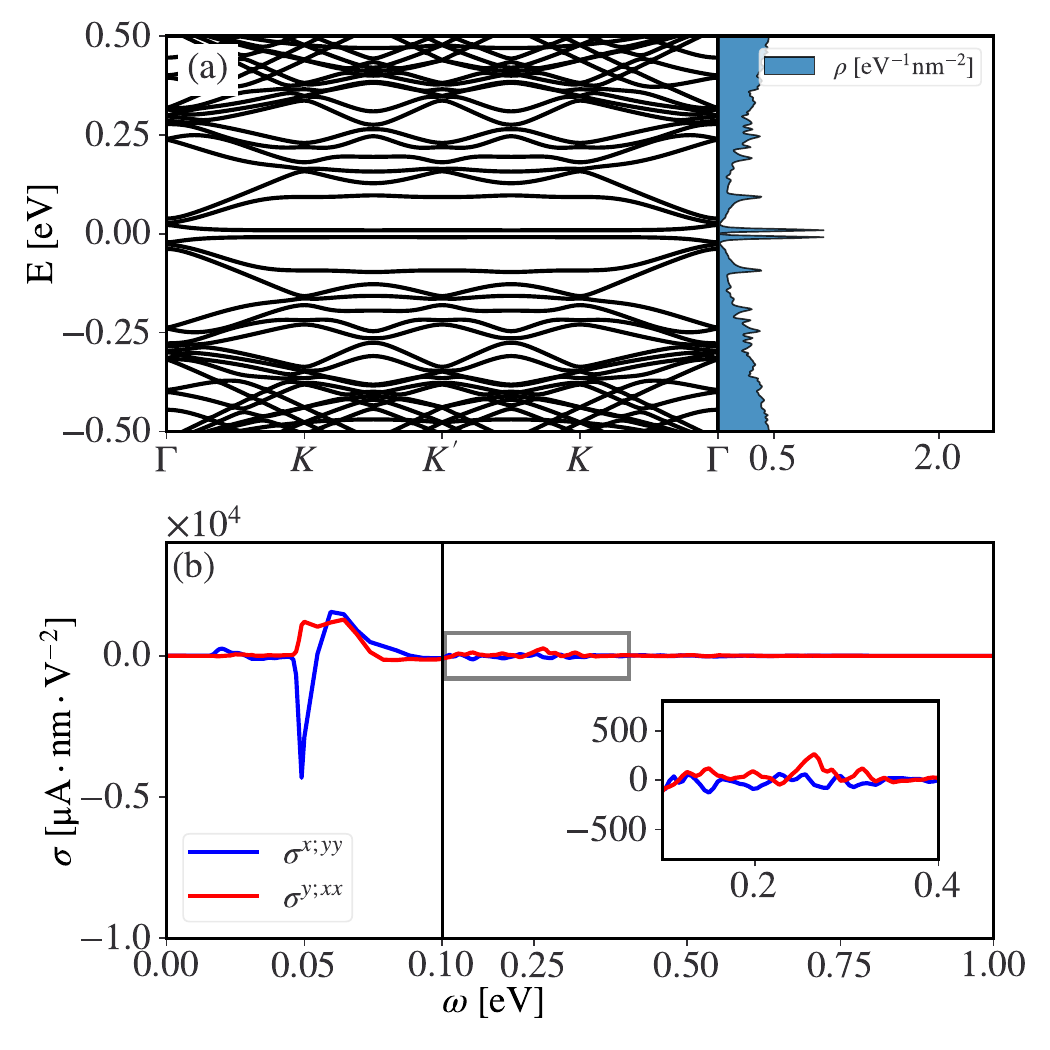}
    \caption{
    (a) Band structure of a MATBG ($\theta=1.09^\circ$) with corrugation factor $r=0.8$, and the associated density of states within the same energy range as in the band structure plot. 
    (b) The shift-current conductivity computed with $T = 0$ and $\mu = 0$ for $\sigma^{x;yy}$ (blue lines) and $\sigma^{y;xx}$ (red lines) as a function of the incident photon energy. The plot zooms in the THz photon energy range below $0.1$ eV. The inset plot zooms in the conductivity plot labeled with the gray rectangle.}
    \label{fig:matbg sc normal}
\end{figure}

In Fig.~\ref{fig:matbg sc normal} we show the SC conductivity coefficients of magic angle twisted bilayer graphene (MATBG) with corrugation $r = 0.8$. The THz far-infrared SC response is still remarkable in normal MATBG, up to orders of $10^3~\mathrm{\mu A \cdot nm \cdot V^{-2}}$, as presented by Fig.~\ref{fig:matbg sc normal}(d). 
The mid-infrared SC response significantly shifts towards the THz regime due to the strong compression of band structure caused by the small twist angle of ($1.09^\circ$).

\begin{figure}[!htbp]
    \centering
    \includegraphics[width=0.95\linewidth]{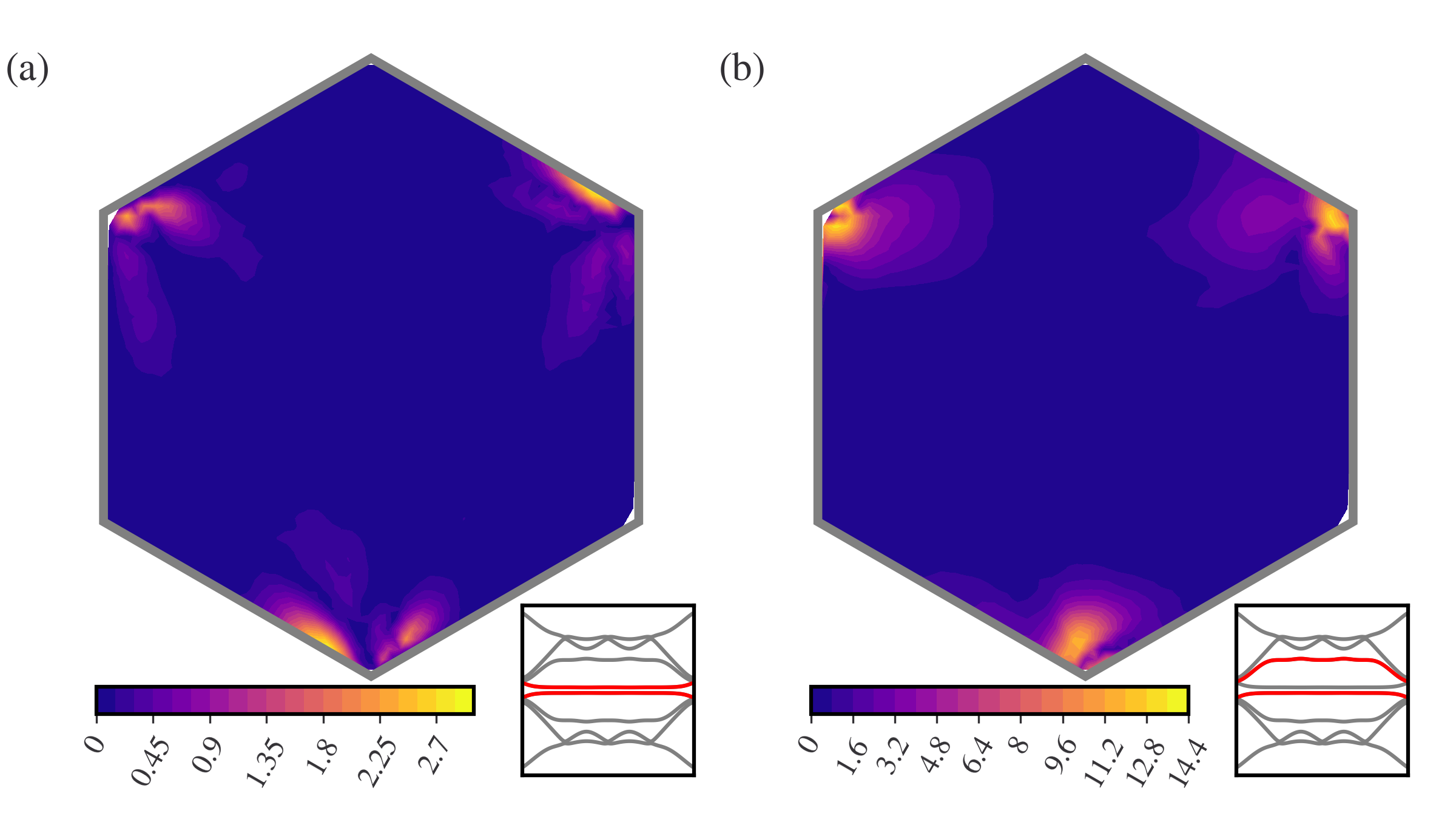}
    \caption{ 
    Dimensionless quantum geometry measure $\tilde{X}^{x;yy}(\bk)$ within the moir\'e Brillouin zone computed for (a) the two middle flattened bands and (b) a middle flat band to dispersive band. Both color plots uses different color scales as indicated by the colorbar beneath each subfigure. The bands contributing to the quantum geometry measure are highlighted with red color in each inset.
    }
    \label{fig:tbg quantum geometry}
\end{figure}

To facilitate the comparison of the quantum geometry measure in hTTG and TBG, the quantum geometry measure of TBG is computed for the two middle bands and and a middle-dispersive band pair in Fig.~\ref{fig:tbg quantum geometry}.

\begin{figure}[!htbp]
    \centering
    \includegraphics[width=0.95\linewidth]{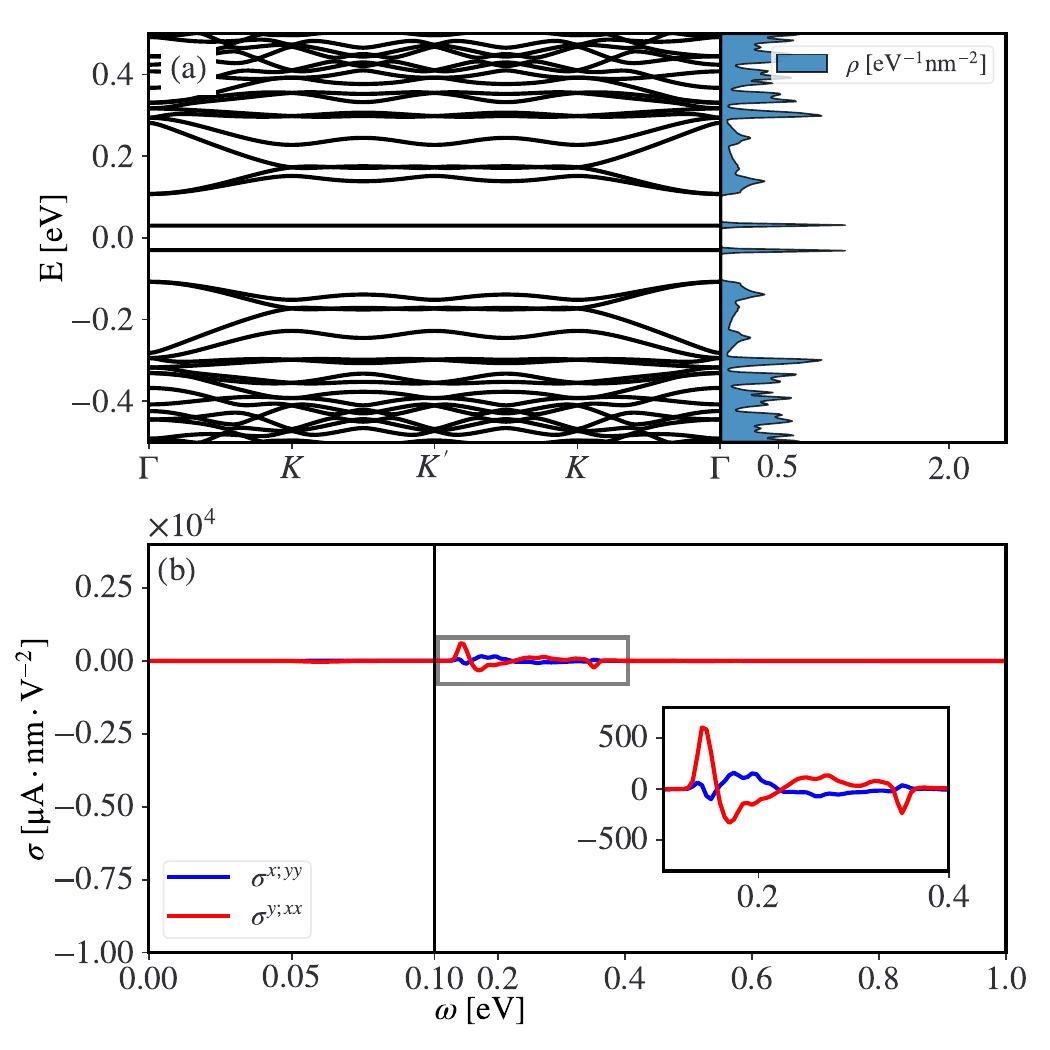}
    \caption{
    (a) and structure of MATBG at the \emph{chiral} limit, and the associated density of states within the same energy range as in the band structure plot. 
    (b) The shift-current conductivity computed with $T = 0$ and $\mu = 0$ for $\sigma^{x;yy}$ (blue lines) and $\sigma^{y;xx}$ (red lines) as a function of the incident photon energy. The plot zooms in the THz photon energy range below $0.1$ eV. The inset plot zooms in the conductivity plot labeled with the gray rectangle.}
    \label{fig:matbg sc chiral}
\end{figure}

The chiral suppression of THz SC response caused by the sublattice polarization of flat-band wave functions is also confirmed in chiral MATBG, as shown in Fig.~\ref{fig:matbg sc chiral}.

\section{Magic angles of ABA-stacked hTTG}\label{app:magic angle}
The magic angle of hTTG is determined by the vanishing dispersion velocity near the Dirac point at $\mathbf{K}$ or $\mathbf{K}'$ of its moir\'e Brillouin zone. The particularity of characterizing the velocity lies in the fact that the Dirac cones in hTTG are in general \emph{anisotropic}, in contrast to TBG where the Dirac cones are always \emph{isotropic}.
\begin{figure}[!htbp]
    \centering
    \includegraphics[width=0.95\linewidth]{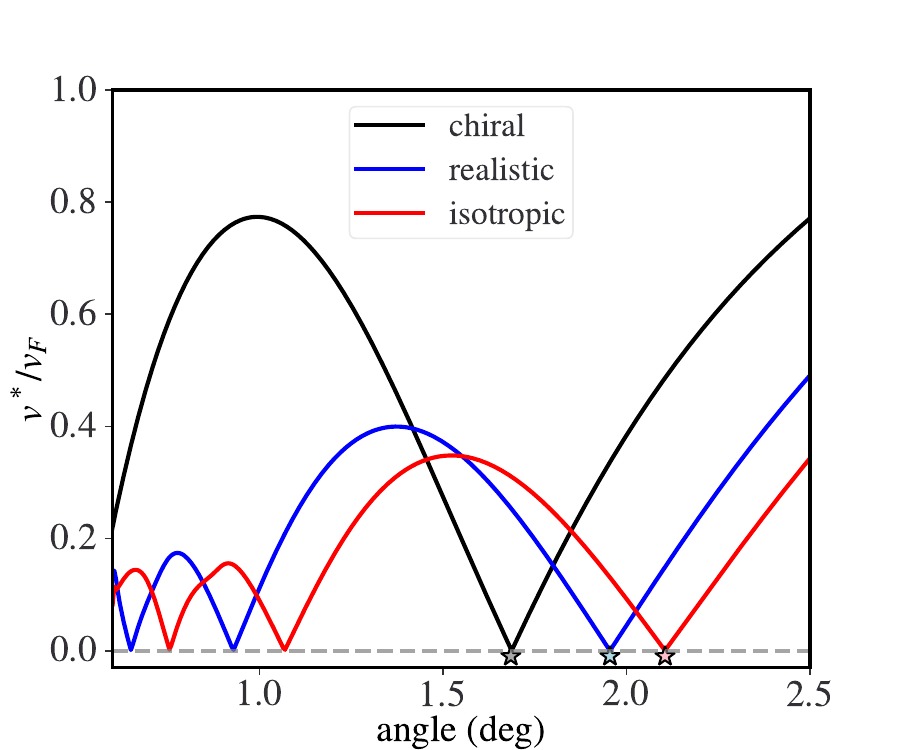}
    \caption{Renormalized Fermi velocity near the $\mathbf{K}$-point as a function of the twist angle. This is computed for the equal twist ABA-stacked hTTG with $r=0$ (chiral limit), $r = 0.8$ (realistic case) and $r = 1$ (isotropic limit), respectively. Magic angles are where the dispersion velocity vanishes. The first magic angle found in the three cases are marked with star symbols.}
    \label{fig:aba vf}
\end{figure}

A singular value decomposition (SVD) technique is employed to characterize the velocity with the two singular values, $s_1$ and $s_2$. Detailed explanations and the mathematical proof of this SVD technique can be found in Ref.~\cite{mao2023supermoire}. Here, we characterize the the dispersion velocity with $v(\theta) = \sqrt{s_1^2 + s_2^2}$, where the results and magic angles are shown in Fig.~\ref{fig:aba vf}.

\clearpage
\bibliography{ref_shift}
\end{document}